\documentstyle[preprint,aps,prb,eqsecnum,version2]{revtex}
\begin{document}
\begin{title}
Chiral two-loop pion-pion scattering parameters from 
crossing-symmetric constraints
\end{title}

\author{G. Wanders\cite{byline}}
\begin{instit}
Institut de physique th\'eorique, Universit\'e de Lausanne,\\ 
CH-1015 Lausanne, Switzerland
\end{instit}
\date{}

\begin{abstract}
Constraints on the parameters in the one- and two-loop pion-pion 
scattering amplitudes of standard chiral perturbation theory are obtained 
from explicitly crossing-symmetric sum rules. These constraints are based on a 
matching of the chiral amplitudes and the physical amplitudes at the symmetry 
point of the Mandelstam plane. The integrals over absorptive parts appearing 
in the sum rules are decomposed into crossing-symmetric low- and high-energy 
components and the chiral parameters are finally related to high-energy 
absorptive parts. A first application uses a simple model of these absorptive 
parts. The sensitivity of the results to the choice of the energy separating 
high and low energies is examined with care. Weak dependence on this energy is 
obtained as long as it stays below $\sim 560$~MeV. Reliable predictions are 
obtained for three two-loop parameters.
\end{abstract}
\begin{pacs}
\end{pacs}

\section{Introduction}

A method for the determination of the parameters appearing in the pion-pion 
scattering amplitudes produced by chiral 
perturbation theory~\cite{Kne1,Bij2} has been proposed in~\cite{Wan3}. 
We demonstrate its practicability in the present work. The method is based on 
sum rules derived from exact analyticity properties combined with crossing 
symmetry. A similar approach has been developed and applied in~\cite{Kne4}. 
Two features distinguish our method: 

(1) Our constraints on the chiral parameters are obtained by matching the true 
amplitudes and their chiral approximations at the symmetry point 
$s=t=u=4M_\pi^2/3$ of the $(s,t,u)$-space ($s$, $t$ and $u$ are the standard 
Mandelstam variables). More precisely we equate coefficients in the Taylor 
expansions of both amplitudes at the symmetry point. 

(2) The Taylor coefficients are expressed by means of crossing-symmetric sum 
rules as integrals over absorptive parts. These integrals are decomposed into 
exactly crossing-symmetric low- and high-energy components. If $\Lambda$ is the 
energy separating low and high energies, we assume that $\Lambda$ can be chosen 
in such a way that the absorptive parts can be approximated by chiral 
absorptive parts below $\Lambda$ and that they are obtained from reliable 
experimental data above $\Lambda$.

The last assumption does not fix $\Lambda$ precisely, whereas the form of the 
conditions constraining the chiral parameters depends explicitly on $\Lambda$. 
Our method is consistent only if these conditions lead to values of the 
parameters depending weakly on $\Lambda$. This stability of the parameters 
with respect to a variation of $\Lambda$ is one of our main concerns. To 
settle this point we work with a simple model of the true absorptive parts 
defined for all energies and at least qualitatively consistent with the data. 
The separating energy $\Lambda$ can be pushed down to the elastic threshold 
$2M_\pi$. We have constructed the constraints for the parameters of the one- 
and two-loop pion-pion amplitudes of standard chiral perturbation theory for 
values of $\Lambda^2$ ranging from $4M_\pi^2$ to $40M_\pi^2$. Our assumptions 
turn out to be valid as long as $4M_\pi^2<\Lambda^2\alt 16M_\pi^2$ 
($=(560~{\rm MeV})^2$).

Before we continue we have to define our notation and the chiral parameters we 
are using. The $s$-channel isospin $I$ amplitude $T^I(s,t,u)$ has the 
partial wave expansion
\begin{equation}
T^I(s,t,u)=\sum_l(2l+1)t_l^I(s)P_l\left(1+{2t\over s-4}\right),\qquad I=0,1,2.
\end{equation}
The pion mass has been set equal to 1, consequently $s+t+u=4$. Below the 
inelastic threshold
\begin{equation}
t_l^I(s)=\sqrt{s\over s-4}\,{\rm e}^{{\rm i}\delta_l^I(s)}\,\sin\,\delta_l(s),
\end{equation}
where the phase shift $\delta_l^I(s)$ is real.

The three amplitudes $T^I$ are obtained from a single function $A(s;t,u)$~:

\begin{eqnarray}
T^0(s,t,u)&=& 3A(s;t,u)+A(t;u,s)+A(u;s,t),\nonumber\\
T^1(s,t,u)&=& A(t;u,s)-A(u;s,t),\\
T^2(s,t,u)&=& A(t;u,s)+A(u;s,t).\nonumber
\end{eqnarray}

In chiral perturbation theory $A$ is the sum of a polynomial $A^{\rm pol}$ and 
an analytic function $A_{\rm cut}$. At 6th order in the momenta and the quark 
masses, each component of $A$ can be written as
\begin{equation}\label{1.4}
A^{(6)}=\lambda^2A_2+\lambda^4A_4+\lambda^6A_6,
\end{equation}
where $\lambda=M_\pi/F_\pi$. In standard chiral perturbation theory we 
have~\cite{Bij2}
\begin{eqnarray}
A_2^{\rm pol}(s)&=&{1\over 32\pi}(s-1), \nonumber\\
A_4^{\rm pol}(s;t,u)&=&{1\over 32\pi}\left[b_1+b_2s+b_3s^2+b_4(t-u)^2\right],
\label{1.5}\\
A_6^{\rm pol}(s;t,u)&=&{1\over 32\pi}\left[b_5s^3+b_6(t-u)^3\right].\nonumber
\end{eqnarray}
The coefficients $b_i$ are the parameters we want to determine. Their 
relationship to the coupling constants in the 4th- and 6th-order effective 
chiral Lagrangian is given in~\cite{Bij2}. We do not use these relations here, 
remarking only that $b_1$, 
$b_2$, $b_3$ and $b_4$ are sums of 4th- and 6th-order terms whereas $b_5$ and 
$b_6$ are 6th-order parameters.

The analytic part of $A^{(6)}$ has the form
\begin{eqnarray}
A_{\rm cut}^{(6)}(s;t,u)&=&{1\over 32\pi}\sum_{\alpha=1}^5\left\{ 
R_\alpha(s)f_\alpha(s)+S_\alpha(t)f_\alpha(t)+S_\alpha(u)f_\alpha(u) \right.
\nonumber\\
&&\label{1.6}\\
&& \qquad \left.+ s\left[T_\alpha(t)f_\alpha(t)+T_\alpha(u)f_\alpha(u)\right]
\right\} \nonumber
\end{eqnarray}
where $R_\alpha(s)$, $S_\alpha(s)$ and $T_\alpha(s)$ are polynomials containing 
4th- and 6th-order terms and $f_\alpha(s)$ are analytic functions with a 
right-hand cut $[4,\infty)$. The polynomials are obtained from formul\ae\ (2) 
and (3) in~\cite{Bij2}: the functions used there have been relabelled 
$f_1(s)=\bar{J}(s)$, $f_\alpha(s)=K_{\alpha-1}(s)$, $\alpha=2,\dots,5$. The 
coefficients of the polynomials in (\ref{1.6}) depend linearily on $b_i$.

Our basic sum rules are obtained from subtracted dispersion relations and 
contain unknown subtraction constants. For this reason the number of useful 
constraints is smaller than the number of parameters, two constraints at 4th 
order and four constraints at 6th order. They are used for the determination 
of $b_3^{(4)}$, $b_4^{(4)}$, $b_3^{(6)}$, $b_4^{(6)}$, $b_5^{(6)}$ and 
$b_6^{(6)}$ at fixed values of $b_1^{(4)}$ and $b_2^{(4)}$. Four parameters 
fulfil the criterion of near $\Lambda$-independence: $b_4^{(4)}$, $b_3^{(6)}$, 
$b_4^{(6)}$ and $b_6^{(6)}$. In spite of the fact that our input modelling the 
experimental data is rather crude, the values we get for the stable 
parameters are entirely compatible with the results derived in~\cite{Kne4}.

The construction of our constraints is outlined in the next two Sections. The 
inputs are defined in Section~IV and the chiral parameters are evaluated in 
Section~V. The results are discussed in Section~VI.

\section{Sum rules for Taylor coefficients at the symmetry point}

The full amplitudes $T^I(s,t,u)$ as well as the chiral amplitudes 
$T_\chi^I(s,t,u)$ are real and regular in the Mandelstam triangle $0< s <4$, 
$0<t<4$, $0<u<4$. This implies that their Taylor expansions at the symmetry 
point $s=t=u=4/3$ converge in a neighbourhood of that point. We assume that 
the $2n$-th order chiral amplitudes approximate the full amplitudes in that 
neighbourhood up to $O\left(p^{2(n+1)}\right)$ corrections. If this assumption 
is correct we can fix the parameters appearing in the chiral amplitudes by 
requiring that $T^I$ and $T_\chi^I$ have identical conveniently truncated 
Taylor expansions at the symmetry point. We adopt this strategy and explore 
its implications.

To write down Taylor expansions for the $T^I$, $s$, $t$ and $u$ 
have to be replaced by two independent variables. Generally the series one 
obtains are constrained by crossing symmetry. It is advisable to resort to a 
procedure which ensures that these constraints are automatically fulfilled. To 
this end the three amplitudes $T^I$ are replaced by three totally symmetric 
amplitudes $G_i$ ($i=0,1,2$) expressed in terms of two new variables 
$x$ and $y$ which are totally symmetric and homogeneous in $s$, $t$ and 
$u$~\cite{Wan5}:
\begin{equation}
x=-{1\over 16}(st+tu+us),\qquad y={1\over 64}stu.
\end{equation}
The $G_i$ have been introduced by Roskies~\cite{Ros6} and are defined as 
follows:
\begin{eqnarray}
G_0(x,y)&=& {1\over 3}\left[T^0(s,t,u)+T^0(u,t,s)+T^0(u,s,t)\right]\nonumber\\
G_1(x,y)&=& {T^1(s,t,u)\over t-u}+{T^1(t,u,s)\over u-s}+{T^1(u,s,t)\over s-t}
\label{eq22}\\
G_2(x,y)&=&{1\over s-t}\left[{T^1(s,t,u)\over t-u}-{T^1(t,s,u)\over s-u}\right]
+ {1\over t-u}\left[{T^1(t,u,s)\over u-s}-{T^1(u,t,s)\over t-s}\right]
\nonumber\nonumber\\
&&+{1\over u-s}\left[{T^1(u,s,t)\over s-t}-{T^1(s,u,t)\over u-t}\right].
\nonumber
\end{eqnarray}

The $T^I$ can be obtained from the $G_i$ (see eq.~(2) in~\cite{Wan3}); their 
symmetry implies crossing symmetry of the $T^I$.

The $G_i$ are regular around the image $(x_s,y_s)$ of the symmetry point in 
the $(x,y)$-space $(x_s=-1/3$, $y_s=1/27$) and have convergent Taylor 
expansions in $x$ and $y$ at that point. They are not constrained by crossing 
symmetry. We equate their first coefficients with the corresponding Taylor 
coefficients of the symmetric chiral amplitudes $G_i^\chi$, the latter being 
given by (\ref{eq22}) with $T^I$ replaced by $
T_\chi^I$.

The Taylor expansions of the amplitudes $G_i^{(2n)}$ obtained from the 
$O(p^{2n})$ amplitudes $T_{(2n)}^I$ are truncated at an order obtained by 
power counting from the definitions (\ref{eq22}). When identifying truncated 
Taylor series at a given order one has to remember that a chiral Taylor 
coefficient gets contributions from all higher orders of the chiral expansion. 
Therefore the conditions one obtains for the $2n$-th order amplitudes hold 
only up to higher order corrections.

For the fourth- and sixth-order amplitudes we get the following constraints:
\begin{mathletters}
\begin{eqnarray}
G_i^{(4)}(x_s,y_s)+O(p^6)&=& G_i(x_s,y_s),\qquad i=0,1,2,\\
{\partial G_0^{(4)}\over \partial x}(x_s,y_s)+O(p^6)&=& {\partial G_0\over 
\partial x}(x_s,y_s),
\end{eqnarray}
\end{mathletters}
\begin{mathletters}
\begin{eqnarray}
G_i^{(6)}(x_s,y_s)+O(p^8)&=& G_i(x_s,y_s),\qquad i=0,1,2,\\
{\partial G_i^{(6)}\over \partial x}(x_s,y_s)+O(p^8)&=& {\partial G_i\over 
\partial x}(x_s,y_s),\qquad i=0,1,\\
{\partial G_0^{(6)}\over \partial y}(x_s,y_s)+O(p^8)&=& {\partial G_0\over 
\partial y}(x_s,y_s).
\end{eqnarray}
\end{mathletters}
The number of constraints obtained in this way at $2n$-th order is equal to 
the number of free parameters appearing in the $2n$-th order chiral amplitudes 
(4 parameters and 4 equations (2.3) at 4th order, 6 parameters and 6 equations 
(2.4) at 6th order). An obvious drawback of conditions (2.3) and (2.4) is 
that, the symmetry point being unphysical, the right-hand sides are not 
directly measurable quantities. Fortunately, there are dispersion relations to 
get rid of this difficulty~\cite{Wan5}. These relations are consequences of 
the exact analyticity properties of the $G_i$ as functions of $x$ and $y$. 
They can be written in the following way:
\begin{equation}
G_i(x,y)= (1-\delta_{i2})G_i(x_0,y_0)+{1\over \pi}\int_4^\infty{\rm d}
\sigma\left[{1\over K(\sigma,x,y)}-{1-\delta_{i2}\over 
K(\sigma,x_0,y_0)}\right]B_i(\sigma,\tau).
\end{equation}
There are once-subtracted dispersion relations for $G_0$ and $G_1$, the 
subtraction being performed at $(x_0,y_0)$, whereas $G_2$ obeys an 
unsubtracted dispersion relation. The integration variable $\sigma$ is an 
energy squared. The denominator function $K(\sigma,x,y)$ has a simple 
expression in terms of the Mandelstam variables:
\begin{equation}
K(\sigma,x,y)=(\sigma-s)(\sigma-t)(\sigma-u) =\sigma^2(\sigma-4)-16\sigma x-
64y,
\end{equation}
if $(s,t,u)$ is a pre-image of $(x,y)$. The second expression results from the 
first one and the definitions (2.1). The $B_i$ are linear combinations of 
$s$-channel absorptive parts $A^I$:
\begin{eqnarray}
B_0(s,t)&=& {1\over 3}(s-t)(2s-4+t)\left[A^0(s,t)+2A^2(s,t)\right],\nonumber\\
B_1(s,t)&=& {1\over 6}(3s-4)\left[2A^0(s,t)-5A^2(s,t)\right]+
\left[{(s-t)(2s-4+t)\over 2t-4+s}-{1\over 2}(2t-4+s)\right]A^1(s,t)\nonumber\\
B_2(s,t)&=& -{1\over 2}\left[2A^0(s,t)-5A^2(s,t)\right]+
{3\over 2}{3s-4\over 2t-4+s}A^1(s,t).
\end{eqnarray}

They have to be evaluated in (2.5) at a $\sigma$-dependent squared momentum 
transfer $\tau$:
\begin{equation}
\tau(\sigma)=-{1\over 2}\left\{(\sigma-4)-\left[(\sigma-4)^2-{16\over 
\sigma+4a}\bigl(a\sigma(\sigma-4)-16(ax_0-y_0)\bigr)\right]^{1\over 
2}\right\}
\end{equation}
where $a$ is the slope of the straight line connecting the point $(x,y)$ to 
the subtraction point $(x_0,y_0)$:
\begin{equation}a={y-y_0\over x-x_0}.
\end{equation}

Details on the derivation of (2.5) are given in~\cite{Wan3}. At fixed 
$(x_0,y_0)$, (2.5) holds true if the slope $a$ belongs to a complex 
neighbourhood of the origin. For our purpose $(x_0,y_0)$ has to be chosen in 
such a way that (2.5) provides a representation of $G_i(x,y)$ in a suitable 
neighbourhood of the symmetry point $(x_s,y_s)$. According to~\cite{Wan3} this 
is the case if $y_0=y_s$ and $-72<x_0<3x_s/2$.

If $\sigma$ is large enough, $\tau(\sigma)$ is real and the point 
$(s=\sigma,t=\tau(\sigma))$ belongs to the physical $s$-channel. If $\sigma$ 
is close to 4, $\tau(\sigma)$ can be complex but the point 
$(s=\sigma,t=\tau(\sigma))$ is inside the large Lehmann ellipse. This implies 
that the absorptive part $A^I(\sigma,\tau)$ is either a physical quantity or is 
obtained from physical partial wave absorptive parts through a convergent 
partial waves expansion. Consequently the dispersion integral in (2.5) is 
itself a physical quantity and this relation produces a representation of 
$G_2(x,y)$ in terms of measurable absorptive parts. This does not hold for 
$G_0(x,y)$ and $G_1(x,y)$ because their values at the subtraction point which 
enter into (2.5) are unphysical with our choice of $(x_0,y_0)$. However, by 
computing derivatives at fixed $(x_0,y_0)$, (2.5) gives expressions for 
$\partial G_i/\partial x$ and $\partial G_i/\partial y$ at $(x_s,y_s)$ which 
do not involve $G_i(x_0,y_0)$. Therefore the dispersion relations provide 
expressions in terms of observable absorptive parts for the right-hand sides 
of eqs~(2.3a, i=2), (2.3b), (2.4a, i=2), (2.4b) and (2.4c).

If one wants to keep eqs~(2.3a) and (2.3b) for $i=0,1$, one can use a 
dispersion relation connecting $G_i(x_0,y_0)$ to the value of $G_i$ at the 
elastic threshold $x=y=0$ (image of $s=4$, $t=u=0$). This value is determined 
by the S-wave scattering lengths, and one obtains expressions for 
$G_i(x_s,y_s)$, $i=0,1$, in terms of physical quantities. However, as the 
experimental scattering lengths are poorly known at present, this would not 
lead to stringent constraints on the chiral parameters. Therefore we restrict 
our discussion to the constraints involving only dispersion integrals over 
absorptive parts.

We end this Section with the expressions of the derivatives $\partial 
G_i/\partial x$, $i=0,1,$ and $\partial G_0/\partial y$ resulting from (2.5):
\begin{eqnarray}
\hspace*{-1cm}{\partial G_i\over \partial x}(x_s,y_s)&=& {32\over \pi}
\int_4^\infty{\rm 
d}\sigma\,{\sigma^2(\sigma-2)+32 y_s\over (K(\sigma,x_s,y_s))^2}\,
{1\over (\sigma-\tau)(2\sigma-4+\tau)}B_i(\sigma,\tau),\qquad i=0,1,\nonumber\\
{\partial G_0\over \partial y}(x_s,y_s)&=& {64\over \pi}\int_4^\infty{\rm 
d}\sigma\,{1\over K(\sigma,x_s,y_s)}\left\{{\sigma(3\sigma-8)-16x_s\over 
K(\sigma,x_s,y_s)}\,{1\over 3}(A^0+2A^2)(\sigma,\tau)\right.\\
&& \quad-\left.{2(\sigma^2(\sigma-2)+64y_s)\over \sigma^{3/2}[\sigma
(\sigma-4)^2-256 y_s]^{1/2}}\,{1\over 3}\left({\partial\over \partial t}
(A^0+2A^2)\right)(\sigma,\tau)\right\}.\nonumber
\end{eqnarray}
In these formulae, $\tau$ is given by (2.8) at $a=0$. As a consequence, the 
integrands do not depend on $x_0$, the position of the subtraction point on 
the line $y=y_s$. This is an advantage of our choice $y_0=y_s$.

\section{Low- and high-energy components}

Our constraints on the chiral pion-pion parameters involve integrals over 
absorptive parts. We have good experimental information in the range 
600~MeV~$<\sqrt{s}<2$~GeV but large uncertainties below this interval. As we 
cannot rely on experimental data in this low energy region we need some  
theoretical input. One way is to exploit the fact that the chiral amplitudes 
are meant to describe low energy scattering correctly. Therefore the low 
energy chiral absorptive parts can be used as approximations of the full 
absorptive parts. This idea was proposed in~\cite{Wan3} and we apply it here. 
A more refined procedure is applied in~\cite{Kne4} where the absorptive parts 
are computed from unitarized chiral S- and P-wave amplitudes.

In a first step the dispersion integral in (2.5) is split into a low energy 
integral $L_i$, from $\sigma=4$ to $\sigma=\Lambda^2$ and a high energy 
integral, $\sigma>\Lambda^2$, $H_i$. The representation (2.5) becomes
\begin{equation}
G_i(x,y)=(1-\delta_{i2})G_i(x_0,y_0)+L_i(x,y)+H_i(x,y).
\end{equation}
This decomposition induces a crossing-symmetric decomposition of the 
amplitudes $T^I$ into low- and high-energy components.

We assume now that $\Lambda^2$ can be chosen in such a way that the $2n$-th 
order chiral absorptive parts $A_\chi^I(\sigma,\tau)$ approximate the full 
absorptive parts for $4\leq\sigma\leq \Lambda^2$ up to $O(p^{2(n+1)})$ 
corrections. Furthermore we have already assumed in Section~II that $G_i$ is 
approximated by $G_i^\chi$ in the neighbourhood of the symmetry point. Thus, 
if $(x,y)$ is close to $(x_s,y_s)$, our assumptions allow us to rewrite 
eq.~(3.1) as follows:
\begin{equation}
G_i^\chi(x,y)=(1-\delta_{i2})G_i(x_0,y_0)+L_i^\chi(x,y)+H_i(x,y) 
+O(p^{2(n+1)}).
\end{equation}
$H_i$ is given by the integral in (2.5) extended over $\sigma>\Lambda^2$; 
$L_i^\chi$ is given by the same integral restricted to 
$4\leq \sigma\leq \Lambda^2$ and with $B_i$ replaced by $B_i^\chi$ .

A key observation is that $G_i^\chi$ and $L_i^\chi$ have the same 
discontinuities across their cuts for $4<\sigma<\Lambda^2$. We show in 
the Appendix that, up to 6th order, this implies that the difference $(G_i^\chi-
L_i^\chi)$ has the following form:
\begin{equation}
G_i^\chi(x,y)-L_i^\chi(x,y)=P_i(x,y)+H_i^\chi(x,y).
\end{equation}
$P_i(x,y)$ is a polynomial of first degree and $H_i^\chi$ is a high-energy 
component given by a dispersion integral starting at $\sigma=\Lambda^2$. 
Equation~(3.2) becomes
\begin{equation}
P_i(x,y)+H_i^\chi(x,y)=(1-\delta_{i2})G_i(x_0,y_0)+H_i(x,y) +O(p^{2(n+1)}).
\end{equation}
The left-hand side is entirely determined by the chiral amplitudes whereas the 
right-hand side is determined, apart from $G_i(x_0,y_0)$, by absorptive parts 
above $\Lambda^2$.

Those equations (2.3) and (2.4) which do not involve $G_i(x_0,y_0)$ are now 
replaced by
\begin{eqnarray}
\left(P_2+H_2^\chi\right)^{(4)}(x_s,y_s)&=& H_2(x_s,y_s)+O(p^6),\nonumber\\
[-2mm]
&& \\[-2mm]
{\partial\over \partial x}\left(P_0+H_0^\chi\right)^{(4)}(x_s,y_s)&=& 
{\partial H_0\over \partial x} (x_s,y_s)+O(p^6).\nonumber
\end{eqnarray}
\begin{eqnarray}
\left(P_2+H_2^\chi\right)^{(6)}(x_s,y_s)&=& H_2(x_s,y_s)+O(p^8),\nonumber\\
{\partial\over \partial x}\left(P_i+H_i^\chi\right)^{(6)}(x_s,y_s)&=& 
{\partial \over \partial x}H_i(x_s,y_s)+O(p^8),\quad i=0,1,\\
{\partial\over \partial y}\left(P_0+H_0^\chi\right)^{(6)}(x_s,y_s)&=& 
{\partial \over \partial y}H_0(x_s,y_s)+O(p^8).\nonumber
\end{eqnarray}

The two equations~(3.5) constrain the four 4th-order parameters and the four 
equations~(3.6) constrain the six 6th-order parameters. These equations 
clearly reduce to eqs~(2.3) and (2.4) if $\Lambda^2=4$. Both sides of the 
equations depend on $\Lambda^2$. However, the restrictions on the chiral 
parameters they imply should not depend on $\Lambda^2$, up to higher-order 
corrections, as long as $\Lambda^2$ is in an interval where our assumption on 
the absorptive parts is valid. Thus we see that eqs~(3.5) and (3.6) provide at 
the same time a tool for the determination of the chiral parameters and a 
verification of the validity of the chiral expansion. We elaborate on these 
two aspects in the following Sections.

\section{Inputs for the parameter constraints}

After these lengthy preliminaries we are ready for a feasibility test of our 
programme. To this end we need an Ansatz for the pion-pion absorptive parts 
and explicit forms for the chiral pion-pion amplitudes. We start with a 
definition of an Ansatz modelling the absorptive parts down to the elastic 
threshold.

For $4<s<51$ (280~MeV~$<\sqrt{s}<1$~GeV) we retain only S- and P-wave 
contributions. A parametrisation proposed by Schenk~\cite{Sch7} is used for 
the corresponding phase shifts:
\begin{equation}
\tan \delta_l^I(s)=\left({1\over 4}(s-4)\right)^l
\left({s-4\over s}\right)^{1\over 2}
{4-s_I\over s-s_I}\left[a_l^I-{1\over 4}(s-4)\tilde{b}_l^I\right],
\end{equation}
where $l=0$ for $I=0,2$ and $l=1$ for $I=1$, $a_l^I$ is a scattering length 
and
\begin{equation}
\tilde{b}_l^I=b_l^I-{4a_l^I\over s_I-4}+(a_l^I)^3,
\end{equation}
where $b_l^I$ is an effective range. In~(4.1), $s_1=M_\rho^2=30.4$ and we 
adopt one of Schenk's choices for the remaining poles: 
$s_0=38=(865~{\rm MeV})^2$, $s_2=-43=-(920~{\rm MeV})^2$. Scattering lengths 
and effective ranges are chosen in accordance with the results of 4th-order 
standard chiral perturbation theory~\cite{Gas8}:
\begin{eqnarray}
a_0^0=0.20,&\qquad&a_0^2=-0.042,\qquad a_1^1=0.037,\nonumber\\[-2mm]
&& \\[-2mm]
b_0^0=0.24,&&b_0^2=-0.075,\qquad b_1^1=0.005.\nonumber
\end{eqnarray}

For $s>51$ we adopt an Ansatz which has been used recently in~\cite{Ana9}. 
In the interval $51<s<110$ (1~GeV~$<\sqrt{s}<1.5$~GeV) we take only the $f_0$ 
contribution
\begin{equation}
A_{f_0}^0(s,t) = 5\pi\Gamma_{f_0}M_{f_0}\sqrt{M_{f_0}^2\over M_{f_0}^2-4}\,P_2
\left(1+{2t\over M_{f_0}^2-4}\right)\delta(s-M_{f_0}^2),
\end{equation}
with $\Gamma_{f_0}=0.896$ ($= 125$~MeV), $M_{f_0}=9.092$ ($= 1269$~MeV).

The high-energy region $s>110$ is described in terms of Pomeron exchange and a 
degenerate $(\rho+f_0)$ Regge trajectory~\cite{Pen10}. If $A^{(I)}(s,t)$ 
designates the isospin $I$ $t$-channel absorptive part, the Pomeron exchange is 
specified by
\begin{equation}
A^{(0)}_{\rm Pom}(s,t)={3x_0\over 32\pi}{1\over M_\pi^2}{\rm e}^{{b\over 
2}t}\left({s\over x_0}\right)^{1+\alpha'_Pt},
\end{equation}
with $x_0=10$~GeV$^2$, $b=10$~GeV$^{-2}$, $\alpha'_P=0.4$~GeV$^{-2}$, 
$A_{\rm Pom}^{(1)}=A_{\rm Pom}^{(2)}=0$~\cite{Bas11}. The $(\rho+f_0)$ exchange 
is represented by the following absorptive parts:
\begin{equation}
A_{\rho,f_0}^{(0)}(s,t)=A_{\rho,f_0}^{(1)}(s,t)=0.6\sin(\pi\alpha(t))\,
\Gamma(1-\alpha(t))\left({s\over 2M_\rho^2}\right)^{\alpha(t)}
\end{equation}
where $\alpha(t)=1/2 +t/(2M_\rho^2)$ and $A_{\rho,f_0}^{(2)}=0$.

The question we are now asking is whether chiral perturbation theory can 
produce amplitudes which are consistent with the absorptive parts defined in 
(4.1), (4.4), (4.5) and (4.6) in the sense that they satisfy the conditions 
(3.5) and (3.6) in a suitable range of values of $\Lambda^2$.

According to eqs (\ref{1.4}-\ref{1.6}) and (2.2) the totally symmetric one- and 
two-loop chiral amplitudes entering into our constraints are of the form
\begin{eqnarray}
G_0^\chi(s,t,u)&=&Q_0(s,t,u)+\sum_{\alpha=1}^5\left[U_\alpha(s)f_\alpha(s)
+U_\alpha(t)f_\alpha(t)+U_\alpha(u)f_\alpha(u)\right], \nonumber\\
G_1^\chi(s,t,u)&=&Q_1(s,t,u)+\sum_{\alpha=1}^5\left\{3\left[T_\alpha(s)f_\alpha
(s)+T_\alpha(t)f_\alpha(t)+T_\alpha(u)f_\alpha(u)\right]\right. \nonumber\\
&& +\left.\left[{1\over s-t}\left(W_\alpha(s)f_\alpha(s)-
W_\alpha(t)f_\alpha(t)\right)+ \mbox{ permutations}\right]\right\},\\
G_2^\chi(s,t,u)&=&Q_2+\sum_{\alpha=1}^5\left\{{1\over s-t}
\left[{1\over t-u}\left(W_\alpha(t)f_\alpha(t)-W_\alpha(u)f_\alpha(u)\right)
\right.\right.\nonumber\\
&& \left.\left.-{1\over u-s}\left(W_\alpha(u)f_\alpha(u)-
W_\alpha(s)f_\alpha(s)\right)\right]+ \mbox{ permutations}\right\}.\nonumber
\end{eqnarray}
In these expressions $Q_0$ and $Q_1$ are polynomials, $Q_2$ is a constant, 
$U_\alpha$, $T_\alpha$ and $W_\alpha$ are polynomials obtained from those in 
(\ref{1.6}) whereas the $f_\alpha$ are the analytic functions appearing there. 
In standard chiral perturbation theory the $Q_i$, $U_\alpha$, $T_\alpha$ 
and $W_\alpha$ depend linearly on the parameters $b_1,\dots,b_6$.

\section{Evaluating 4th- and 6th-order parameters}

The ingredients collected in the last Section allow us to compute the left- 
and right-hand sides of eqs~(3.5) and (3.6). The 4th-order equations (3.5) 
involve only two parameters, $b_3$ and $b_4$:
\begin{eqnarray}
{3\over 32\pi}\lambda^4\left[-b_3+3b_4+C_1(\Lambda^2)\right]^{(4)}&=& 
H_2(\Lambda^2)+O(p^6),\nonumber\\[-2mm]
&& \\[-2mm]
{1\over \pi}\lambda^4\left[b_3+3b_4+C_2(\Lambda^2)\right]^{(4)}&=& 
{\partial H_0\over \partial x}(\Lambda^2)+O(p^6). \nonumber
\end{eqnarray}

The $\Lambda^2$ dependencies are indicated explicitly and every reference to 
the symmetry point is dropped. For instance $H_2(\Lambda^2)$ is the 
$\Lambda^2$-dependent value of $H_2(x_s,y_s)$ and $C_1(\Lambda^2)$ is the term 
of $(P_2+H_2^\chi)^{(4)}(x_s,y_s)$ which is does not depend on the $b_i$.

For the time being it is convenient to work with the combinations 
$(-b_3+3b_4)$ and $(b_3+3b_4)$ rather than $b_3$ and $b_4$. Figure~1 displays 
the values obtained form eqs~(5.1) ignoring the $O(p^6)$ corrections. The 
variations of $H_2$ and $\partial_xH_0$ as functions of $\Lambda^2$ are 
compensated to a large extent by the $\Lambda^2$-dependence of $C_1$ and 
$C_2$, but a sizeable $\Lambda^2$-dependence remains. One may ask if it could 
be compensated by $\Lambda^2$-dependent $O(p^6)$ corrections. It is delicate 
to guess the order of magnitude of such corrections but we 
do not have to speculate at this point: as we shall see, the $O(p^6)$ 
corrections we are looking for are obtained from the 6th-order equations 
(3.6).

Two of the equations (3.6) are corrected versions of the 4th-order eqs~(5.1). 
They have the form:
\begin{eqnarray}
\lambda^4\left(-b_3+3b_4+C_1(\Lambda^2)\right)^{(6)}\!+\lambda^6
\left[-4(b_5-
b_6)+g_1(b_1,b_2,b_3,b_4,\Lambda^2)^{(4)}\right]\!&=&
{32\pi\over 3}H_2(\Lambda^2)+O(p^8),\nonumber\\
&& \\
\lambda^4\left(b_3+3b_4+C_2(\Lambda^2)\right)^{(6)}\!+\lambda^6
\left[6b_5-3b_6+g_2(b_1,b_2,b_3,b_4,\Lambda^2)^{(4)}\right]\!&=& 
\pi{\partial H_0\over \partial x}(\Lambda^2)+O(p^8). \nonumber
\end{eqnarray}

In these equations
\begin{equation}
b_i^{(6)}=b_i^{(4)}+\lambda^2\delta b_i,\quad i=3,4,\qquad
C_k^{(6)}=C_k^{(4)}+\lambda^2\delta C_k,\quad k=1,2,
\end{equation}
where $\delta b_i$ and $\delta C_k$ are 6th-order contributions and 
$\delta C_k$ is known.

In addition to (5.3) we have two true 6th-order equations:
\begin{eqnarray}
\lambda^4C_3^{(6)}(\Lambda^2)+\lambda^6
\left[b_5-3b_6+g_3(b_1,b_2,b_3,b_4,\Lambda^2)^{(4)}\right]&=& 
{\pi\over 6}{\partial H_0\over \partial y}(\Lambda^2)+O(p^8),\nonumber\\[-2mm]
&& \\[-2mm]
\lambda^4C_4^{(6)}(\Lambda^2)+\lambda^6
\left[3(b_5+b_6)+g_4(b_1,b_2,b_3,b_4,\Lambda^2)^{(4)}\right]&=& 
2\pi{\partial H_1\over \partial x}(\Lambda^2)+O(p^8). \nonumber
\end{eqnarray}
The $g_k$ in (5.2) and (5.4) are known linear and homogeneous combinations of 
$b_1$, $b_2$, $b_3$ and $b_4$ with $\Lambda^2$-dependent coefficients. 
Fourth-order values have to be inserted.

The constants $C_3$ and $C_4$ are of the form indicated in (5.3): they are 
sums of 4th-order terms and 6th-order corrections. We meet here a peculiarity 
of our programme. Although equations (5.4) do not appear at 4th-order level 
because they come from $O(p^6)$ terms in the expansions of the $T_\chi^I$ at 
the symmetry point, genuine 4th-order terms enter into these equations. This 
happens because each coefficient in the Taylor series of $H_i^\chi$ has an 
expansion starting with a 4th-order term, in contrast with $\partial 
P_0/\partial y$ and $\partial P_1/\partial x$ which contain no such terms. 
In practice, the presence of 4th-order contributions in eqs~(5.4) is harmless 
because these contributions are small, especially for large enough 
$\Lambda^2$, so that they have effectively the same order of magnitude as the 
true 6th-order terms.

Eliminating the two 6th-order parameters $b_5$ and $b_6$ from eqs~(5.2) and 
(5.4) gives two equations relating $b_1$, $b_2$, $b_3$ and $b_4$:
\begin{eqnarray}
\lambda^4\left(-b_3+3b_4+C_5(\Lambda^2)\right)^{(6)}&+&\lambda^6
g_5(b_1,b_2,b_3,b_4,\Lambda^2)^{(4)}\nonumber\\
&=&{\pi\over 3}\left[32 H_2(\Lambda^2)+{\partial H_0\over \partial y}(\Lambda^2)
+12{\partial H_1\over \partial x}(\Lambda^2)\right]+O(p^8),\nonumber\\[-2mm]
&& \\[-2mm]
\lambda^4\left(b_3+3b_4+C_6(\Lambda^2)\right)^{(6)}&+&\lambda^6
g_6(b_1,b_2,b_3,b_4,\Lambda^2)^{(4)}\nonumber\\
&=&{\pi\over 3}\left[3{\partial H_0\over \partial x}(\Lambda^2)
-{\partial H_0\over \partial y}(\Lambda^2)
-12{\partial H_1\over \partial x}(\Lambda^2)\right]+O(p^8). \nonumber
\end{eqnarray}

As the 4th-order constraints (5.1) do not provide well defined values for 
$b_3^{(4)}$ and $b_4^{(4)}$ we insert $b_3^{(6)}$ and $b_4^{(6)}$ into $g_5$ 
and $g_6$. This amounts to a redefinition of the $O(p^8)$ corrections. We 
retain $b_1^{(4)}$ and $b_2^{(4)}$ and drop the indices labelling orders in 
the chiral expansion from now on. To illustrate the quantitative content of 
eqs~(5.5), we write them explicitly for $\Lambda^2=10$, without mention of the 
$O(p^8)$ corrections:
\begin{eqnarray}
-(1-[0.14])b_3+(3+[0.24])b_4+4.17\cdot 10^{-4}\qquad\qquad && \nonumber\\
-[6.12\cdot 10^{-3}]b_1-[4.26\cdot 10^{-3}]b_2+[4.78\cdot 10^{-4}]&=& 
1.81\cdot 10^{-2}\nonumber\\[-2mm]
&&\\[-2mm]
(1-[0.12])b_3+(3+[0.10])b_4-3.18\cdot 10^{-3}\qquad\qquad&& \nonumber\\
+[1.39\cdot 10^{-3}]b_1-[1.29\cdot 10^{-2}]b_2-[4.56\cdot 10^{-4}]&=& 
7.28\cdot 10^{-3}.\nonumber
\end{eqnarray}

Both sides of eqs~(5.5) have been divided by $\lambda^4$. The square brackets 
indicate two-loop terms: numbers without square brackets are 4th-order terms.

For $\Lambda^2=10$, the 4th-order version of eqs~(5.6), i.e.~eqs~(5.1), is
\begin{eqnarray}
-b_3+3b_4-1.52\cdot 10^{-3}&=& 1.54\cdot 10^{-2}\nonumber\\[-2mm]
&&\\[-2mm]
b_3+3b_4-6.50\cdot 10^{-4}&=& 1.15\cdot 10^{-2}.\nonumber
\end{eqnarray} 
These equations are not identical to eqs~(5.6) with the square brackets 
dropped. This is due to the 4th-order contributions of $\partial 
H_0^\chi/\partial y$ and $\partial H_1^\chi/\partial x$, mentioned previously, 
in eqs~(5.6).

Equations (5.5) are treated as determining $b_3$ and $b_4$ in terms of $b_1$ 
and $b_2$, using the values of these parameters obtained from 1-loop 
standard perturbation theory~\cite{Gas12,Bij14}:
\begin{eqnarray}
b_1&=& {1\over 96\pi^2}\left[8\bar{l}_1-3\bar{l}_3-12\bar{l}_4+{13\over 
3}\right]\nonumber\\[-2mm]
&&\\[-2mm]
b_2&=& {1\over 24\pi^2}\left[-2\bar{l}_1+3\bar{l}_4-{1\over 3}\right].\nonumber
\end{eqnarray}
With the central values~\cite{Gas12,Bij14} $\bar{l}_1=-1.7$, $\bar{l}_3=2.9$, 
$\bar{l}_4=4.6$ we get
\begin{equation}
b_1=-7.7\cdot 10^{-2},\qquad b_2=7.1\cdot 10^{-2}.
\end{equation}
Inserting these values into the $\Lambda^2=10$ equations (5.6) gives
\begin{equation}
(-b_3+3b_4)^{(6)}=1.63\cdot 10^{-2},\qquad (b_3+3b_4)^{(6)}=1.12\cdot 10^{-2}.
\end{equation}

The corresponding 4th-order values are $1.70 \cdot 10^{-2}$ and $1.21\cdot 
10^{-2}$, quite close to the 6th-order values~(5.10). The variation of 
$(\mp b_3+3b_4)$ at 4th and 6th order as functions of $\Lambda^2$ 
obtained from eqs~(5.1) and (5.5) is shown in Fig.~1. The striking feature is 
that $(\mp b_3+3b_4)^{(6)}$ is independent of $\Lambda^2$ within $4\%$ for 
$4<\Lambda^2<16$. On the other hand, the difference between the 6th- and 
4th-order values can be substantial, especially for $(b_3+3b_4)$ and 
$\Lambda^2$ between 4 and 10.

The values of $b_3$ and $b_4$ as obtained from $(-b_3+3b_4)$ and $(b_3+3b_4)$ 
are displayed in Fig.~2: $b_4^{(4)}$ and $b_4^{(6)}$ are practically constant, 
their difference due to a 6th-order correction is of the order of $4\%$. The 
parameter $b_3$ is not as robust as $b_4$. The way it is obtained amplifies 
the uncertainties. At 4th order, they must have the same magnitude as the 
computed values. The situation improves at 6th order and $b_3$ becomes 
independent of $\Lambda^2$ between 4 and 16 within $6\%$.

Ignoring all uncertainties in the input quantities $H_i$ and taking the 
variations with $\Lambda^2$ for $4<\Lambda^2<16$ into account leads to the 
following 6th-order values of $b_3$ and $b_4$~:
\begin{equation}
b_3=(-2.55\pm 0.20)\cdot 10^{-3},\qquad b_4=(4.55\pm 0.15)\cdot 10^{-3}.
\end{equation}
The only clear 4th-order result is $b_4^{(4)}=(4.85\pm 0.15)\cdot 10^{-3}$.

Once $b_3$ and $b_4$ are known, eqs~(5.4) determine $b_5$ and $b_6$. With (5.9) 
and the central values in (5.11), this leads to $b_6$ depending 
weakly on $\Lambda^2$, as shown in Fig.~3. In contrast, $b_5$ exhibits a strong 
$\Lambda^2$-dependence: it 
decreases from $1.4\cdot 10^{-4}$ to $0.66\cdot 10^{-4}$ as $\Lambda^2$ varies 
between 4 and 16. Comparing Fig.~2 and Fig.~3, we observe that $b_4^{(4)}$ and 
$b_6^{(6)}$ behave similarly, and so do $b_3^{(4)}$ and $b_5^{(6)}$. This is 
possibly due to the similar roles which the pairs $(b_3,b_4)$ and $(b_5,b_6)$ 
play in $A^{\rm pol}(s;t,u)$ (eq.~(\ref{1.5})). Furthermore, the 4th-order 
absorptive parts do not depend on $(b_3,b_4)$ and the 6th-order absorptive 
parts do not depend on $(b_5,b_6)$. One may expect that $b_5$ would become 
stable at 8th order in the same way as $b_3$ is stabilized at 6th order.

Ignoring again the uncertainties in the $H_i$, Fig.~3 gives
\begin{equation}
b_6=(9.2\pm 0.3)\cdot 10^{-5}.
\end{equation}
The values of $b_3$, $b_4$ and $b_6$ indicated in (5.11) and (5.12) are 
compatible with the results obtained in~\cite{Kne4}. Concerning $b_5$, no 
large instability shows up in~\cite{Kne4}. This is probably related to the use 
of unitarized absorptive parts. In any case the order of magnitude we obtain 
is the same as in~\cite{Kne4}.

\section{Discussion}

The last Section provides evidence that our programme is working. It produces 
three 6th-order parameters which are stable in the sense that they depend 
weakly on $\Lambda$, the energy separating low and high energies as long as 
$2<\Lambda\alt 4$. The following comments summarise the salient features of 
our findings.

1. The fact that $b_3$, $b_4$ and $b_6$ are nearly independent of 
$\Lambda^2$ at 6th order is not the result of a weak $\Lambda^2$-dependence of 
the individual terms in eqs.~(5.2) and (5.4). In fact they vary strongly with 
$\Lambda^2$: for instance $H_2(\Lambda^2)$ in eq.~(5.2) increases from 
$1.01\cdot 10^{-2}$ to $1.69\cdot 10^{-2}$ when $\Lambda^2$ goes from 4 to 16. 
Very delicate compensations are at work and are responsible for practically 
constant 6th-order parameters.

2. The choice of the values (5.8) for $b_1$ and $b_2$ plays a role in the 
constancy of $b_3^{(6)}$, $b_4^{(6)}$ and $b_6^{(6)}$. Substantially different 
values would destroy the compensations mentioned under 1. In fact, the 
constancy of $b_3^{(6)}$ and $b_4^{(6)}$ at the level displayed in Fig.~2 is 
certainly lost if $(b_1,b_2)$ is outside the square $(-2\cdot 10^{-1},\,0)
\times (0,\,2\cdot 10^{-1})$.

3. One of our basic assumptions is that low-energy absorptive parts can be 
approximated by chiral absorptive parts. To check this we have computed the 
4th- and 6th-order S- and P-wave absorptive parts. The 4th-order absorptive 
parts do not depend on the $b_i$: the central values (5.9) and (5.11) have been 
used in the 6th-order calculations. Figure~4 shows that in the case of 
$\mbox{\rm Im } f_0^0(s)$ there is good agreement between the 6th-order chiral 
absorptive part and our Ansatz (4.1) if $4<s<18$. There is also a spectacular 
improvement with respect to the 4th-order result. The agreement is not as good 
for $\mbox{\rm Im } f_1^1(s)$ and $\mbox{\rm Im } f_0^2(s)$. This does not 
affect our results because these absorptive parts are small at low energies and 
the contributions of $\mbox{\rm Im } f_0^0(s)$ dominate.

4. According to eqs~(5.5), $b_3$ and $b_4$ depend additively on $C_1$, $C_2$, 
$b_2$, $H_2$, $\partial H_0/\partial x$, $\partial H_1/\partial x$ and 
$\partial H_0/\partial y$.This means that their values are dictated not only 
by the inputs $H_2$, $\partial H_0/\partial x,\; \dots$ but also by the 
structure of the chiral amplitudes. The inputs produce contributions given in 
Table~1 for four values of $\Lambda^2$. This table also displays the relative 
sizes of the contributions coming from the three energy intervals 
$(\Lambda,\,1~{\rm GeV})$, 
$(1~{\rm GeV},\,1.5~{\rm GeV})$ and $(1.5~{\rm GeV},\,\infty)$ appearing in the 
input defined in Section~IV. In this table, $b_3[s_1,s_2]$ and $b_4[s_1,s_2]$ 
are the contributions of the $H_i$ to $b_3^{(6)}$ and $b_4^{(6)}$ coming from 
integrals restricted to $s_1<\sigma<s_2$.

We are working here with a crude Ansatz for the absorptive parts and we are 
primarily interested in estimates of the $b_i$ rather than precise 
determinations including error evaluations. Table~1 indicates how our input 
should be improved to favour a precise determination of $b_3$ and $b_4$. 
Although the major parts of $b_3[\Lambda^2,\infty]$ and $b_4[\Lambda^2,\infty]$ 
come from energies below 1~GeV, the energies above 1~GeV contribute to 
$b_4[\Lambda^2,\infty]$ at the 10\%\ level and there is a strong cancellation 
at work for $b_3[\Lambda^2,\infty]$. We use a Regge Ansatz on 
$(1.5\mbox{ GeV},\infty)$ and it would be difficult to estimate the errors 
associated with this. The cut at $s=110$ ($=(1.5~{\rm GeV})^2$) should 
be replaced by a cut at $s= 200$ ($=(2~{\rm GeV})^2$), experimental 
data used on $(51,200)$, the Regge Ansatz being restricted to $(200,\infty)$. 
This would approximately halve the Regge contribution.

5. The implications of our findings for the 6th-order coupling constants are 
beyond the scope of the present work. We observe only that the 4th-order 
relation
\begin{equation}
\bar{l}_2=96\cdot\pi^2\cdot b_4^{(4)}+{5\over6}
\end{equation}
allow a determination of $\bar{l}_2$ (notice that $\bar{l}_2$ does not enter 
into our evaluation of $b_1$ and $b_2$, eqs~(5.8)). This gives $\bar{l}_2=5.4$ 
which is below the central value $\bar{l}_2=6.1$ obtained from the analysis of 
$K_{l4}$ decays~\cite{Bij14}. A low value such as this has also been extracted 
from pion-pion scattering in~\cite{Ana9} and \cite{Pen15}. Still at 4th order 
we have
\begin{equation}
b_3^{(4)}={1\over 96\pi^2}\left[2\bar{l}_1+\bar{l}_2-{7\over 3}\right].
\end{equation}
With the value $\bar{l}_1=-1.7$ used in eqs~(5.8), and $\bar{l}_2=5.4$, this 
gives $b_3^{(4)}=-1.6\cdot 10^{-3}$. This is compatible with the range of 
variation of $b_3^{(4)}$ displayed in Fig.~2 and confirms the credibility of 
our approach.

\acknowledgments

Discussions with P.~B\"uttiker, J.~Gasser, H.~Leutwyler, J.~Stern and 
D.~Toublan are gratefully acknowledged. D.~Toublan was helpful in the 
preparation of the figures. 

\setcounter{equation}{0}
\unletteredappendix{Determination of low-energy polynomials}

We have to show that the terms $P_i$ in eq.~(3.3) are polynomials. We do this 
in detail for $P_0$.

The functions $f_\alpha$ vanish at the origin and can be obtained from a 
once-subtracted dispersion relation:
\begin{equation}
f_\alpha(s)={s\over \pi}\int_4^\infty{{\rm d}\sigma\over \sigma}{1\over 
\sigma-s}\mbox{\rm Im } f_\alpha(\sigma).
\end{equation}
By writing the right-hand side as the sum of a low-energy and a high-energy 
integral, $f_\alpha$ is decomposed into a low- and a high-energy component. 
Introducing this decomposition into the expression (4.7) of $G_0$ gives
\begin{equation}\label{A.2}
G_0^\chi(s,t,u)=Q_0(s,t,u)+{1\over \pi}\sum_\alpha\int_4^{\Lambda^2}{{\rm 
d}\sigma\over \sigma}\left[{sU_\alpha(s)\over \sigma-s}+{tU_\alpha(t)\over 
\sigma-t}+{uU_\alpha(u)\over \sigma-
u}\right]\mbox{\rm Im } f_\alpha(\sigma)+H_0^\chi(s,t,u). 
\end{equation}

The high-energy component $H_0^\chi$ is the sum of integrals in (\ref{A.2}) 
extended to $[\Lambda^2,\infty)$. The integrals in (\ref{A.2}) can be rewritten 
as 
\begin{eqnarray}
\lefteqn{-{1\over\pi}\int_4^{\Lambda^2}{{\rm d}\sigma\over \sigma}\left[
U_\alpha(s,\sigma)+U_\alpha(t,\sigma)+U_\alpha(u,\sigma)\right]
\mbox{\rm Im } f_\alpha(\sigma)} &&\nonumber\\
&&\qquad +{1\over\pi}\int_4^{\Lambda^2}{{\rm d}\sigma\over \sigma}\left[
{1\over\sigma-s}+{1\over\sigma-t}+{1\over\sigma-
u}\right]U_\alpha(\sigma)\mbox{\rm Im } f_\alpha(\sigma),\label{A.3}
\end{eqnarray}
where
\begin{equation}
U_\alpha(s,\sigma)={sU_\alpha(s)-\sigma U_\alpha(\sigma)\over s-\sigma}
\end{equation}
is a polynomial. On the other hand, (4.7) and (2.7) tell us that 
$\displaystyle B_0^\chi=\sum_\alpha B_{0,\alpha}^\chi$ with
\begin{equation}
B_{0,\alpha}^\chi(s,t)=(s-t)(2s-4+t)U_\alpha(s)\mbox{\rm Im } f_\alpha(s).
\end{equation}
According to (2.5) and (2.6), we have $L_0^\chi=\sum_\alpha 
L_{0,\alpha}^\chi$, where
\begin{equation}
\hspace*{-6mm} L_{0,\alpha}^\chi(s,t,u)={1\over \pi}\int_4^{\Lambda^2}{\rm d}
\sigma\left[{1\over(\sigma-s)(\sigma-t)(\sigma-u)}
-{1\over(\sigma-s_0)(\sigma-t_0)(\sigma-
u_0)}\right]B_{0,\alpha}^\chi(\sigma,\tau).
\end{equation}

It turns out that the difference between the second integral in (\ref{A.3}) and 
$L_{0,\alpha}^\chi$ is equal to this integral with $(s,t,u)$ replaced by 
$(s_0,t_0,u_0)$. Consequently (3.3) is true for $i=0$ with
\begin{eqnarray}
P_0(s,t,u)&=&Q_0(s,t,u)-{1\over\pi}\sum_\alpha
\int_4^{\Lambda^2}{{\rm d}\sigma\over \sigma}\left[
U_\alpha(s,\sigma)+U_\alpha(t,\sigma)+U_\alpha(u,\sigma)\right]
\mbox{\rm Im } f_\alpha(\sigma)\nonumber\\
&&\qquad +{1\over\pi}\int_4^{\Lambda^2}{\rm d}\sigma\left[
{1\over\sigma-s}+{1\over\sigma-t}+{1\over\sigma-
u}\right]U_\alpha(\sigma)\mbox{\rm Im } f_\alpha(\sigma). \label{A.4}
\end{eqnarray}

In terms of $x$ and $y$, and at fixed $(x_0,y_0)$, $P_0(x,y)$ is a polynomial 
of first degree.

Similarly,
\begin{eqnarray}
P_1(s,t,u)&=&Q_1(s,t,u)-{1\over\pi}\sum_\alpha
\int_4^{\Lambda^2}{{\rm d}\sigma\over \sigma}\left\{
3\left[T_\alpha(s,\sigma)+T_\alpha(t,\sigma)+T_\alpha(u,\sigma)\right]\right.
\nonumber\\
&&\qquad +\left.\left[{1\over s-t}\left(W_\alpha(s,\sigma)-
W_\alpha(t,\sigma)\right)+\mbox{ permutations}\right]\right\} +
\mbox{ const.}\\[4mm] 
P_2&=&Q_2-{1\over\pi}\sum_\alpha
\int_4^{\Lambda^2}{{\rm d}\sigma\over \sigma}\left\{{1\over s-t}\left[{1\over 
t-u}\left(W_\alpha(t,\sigma)-W_\alpha(u,\sigma)\right)\right.\right.\nonumber\\
&&\qquad -\left.\left.{1\over u-s}\left(W_\alpha(u,\sigma)-
W_\alpha(s,\sigma)\right)\right]+\mbox{ permutations}\right\}.
\end{eqnarray}

The polynomials $W_\alpha(s,\sigma)$ and $T_\alpha(s,\sigma)$ are obtained in 
the same way as $U_\alpha(s,\sigma)$ in (\ref{A.4}). It turns out that 
$P_1(x,y)$ is linear in $x$, independent of $y$ and $P_2$ is a constant.

\figure{The variations of the combinations $(\pm b_3+3b_4)$ as functions 
of $\Lambda^2$. The 4th-order values (dashed lines) are obtained from 
eqs~(5.1) and the 6th-order values (full lines) come from eqs~(5.5) with 
$b_1=-7.7\cdot 10^{-2}$, $b_2=7.1\cdot 10^{-2}$.}

\figure{The values of $b_3$ and $b_4$ at 4th order (dashed lines) and 
6th order (full lines) resulting from Fig.~1.}

\figure{The $\Lambda^2$-dependent values of $b_5$ and $b_6$ obtained 
from eqs~(5.4) with $b_1=-7.7\cdot 10^{-2}$, $b_2=7.1\cdot 10^{-2}$, 
$b_3=-2.55\cdot 10^{-3}$ and $b_4=4.55\cdot 10^{-3}$.}

\figure{Comparison of the S- and P-wave absorptive parts defined by the 
Ansatz~(4.1) (full lines) with the chiral 4th-order (dashed lines) and 
6th-order (dashed-dotted lines) absorptive parts. The latter are obtained from 
eq.~(1.6) with the same values of $b_1$, $b_2$, $b_3$ and $b_4$ as in 
Fig.~3.}

\renewcommand{\arraystretch}{1.2}
\begin{table}
\caption{The $\Lambda^2$-dependent contributions to $b_3$ and $b_4$ at 
6th order coming from the input quantities $H_2$, $\partial H_0/\partial x$, 
$\partial H_1/\partial x$ and $\partial H_0/\partial y$ according to 
eqs~(5.5). The full contributions are given in the second column. The relative 
sizes of the contributions of the three energy intervals $\Lambda^2<s<51$, 
$51<s<110$ and $s>110$ appear in columns 3-5.}

\vspace*{8mm}
\begin{tabular}{rrrrr}
$\Lambda^2$ &$b_3\left[\Lambda^2,\infty\right]$ & 
$\displaystyle {b_3\left[\Lambda^2,51\right]\over b_3\left[\Lambda^2,\infty
\right]}$ &
$\displaystyle {b_3\left[51,110\right]\over b_3\left[\Lambda^2,\infty\right]}$ &
$\displaystyle {b_3\left[110,\infty\right]\over 
b_3\left[\Lambda^2,\infty\right]}$ \\[2mm]
\tableline
$4$ & $-5.94\cdot 10^{-3}$ & $1.011$ & $-0.060$ & $0.049$\\
$8$ & $-5.78\cdot 10^{-3}$ & $1.004$ & $-0.060$ & $0.055$\\
$12$ & $-7.60\cdot 10^{-3}$ & $1.004$ & $-0.049$ & $0.045$\\
$16$ & $-8.74\cdot 10^{-3}$ & $1.007$ & $-0.046$ & $0.038$\\
\tableline
\tableline
$\Lambda^2$ & $b_4\left[\Lambda^2,\infty\right]$ & 
$\displaystyle {b_4\left[\Lambda^2,51\right]\over b_4\left[\Lambda^2,\infty
\right]}$ &
$\displaystyle {b_4\left[51,110\right]\over b_4\left[\Lambda^2,\infty\right]}$ &
$\displaystyle {b_4\left[110,\infty\right]\over 
b_4\left[\Lambda^2,\infty\right]}$ \\[2mm]
\tableline
$4$ & $4.94\cdot 10^{-3}$ & $0.905$ & $0.017$ & $0.078$\\
$8$ & $4.49\cdot 10^{-3}$ & $0.900$ & $0.018$ & $0.082$\\
$12$ & $4.02\cdot 10^{-3}$ & $0.894$ & $0.019$ & $0.088$\\
$16$ & $3.37\cdot 10^{-3}$ & $0.888$ & $0.019$ & $0.093$\\
\end{tabular}
\end{table}
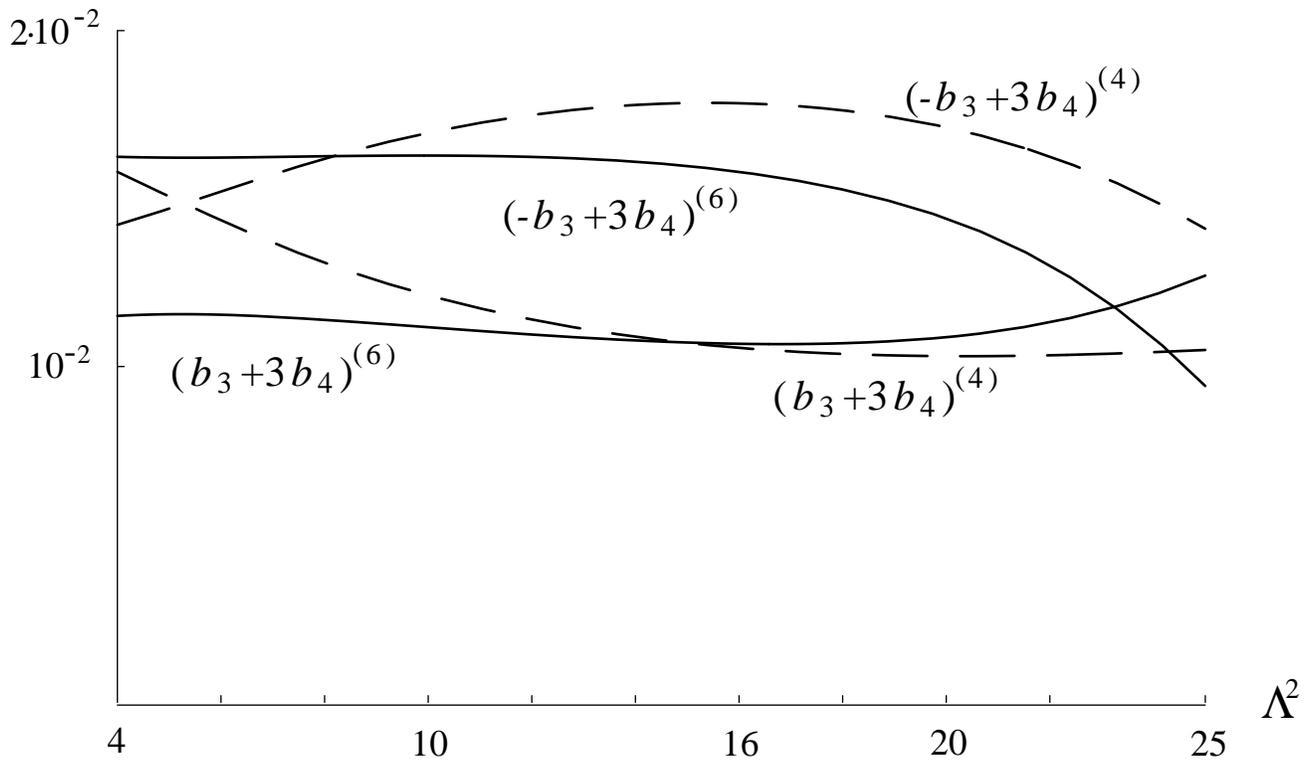
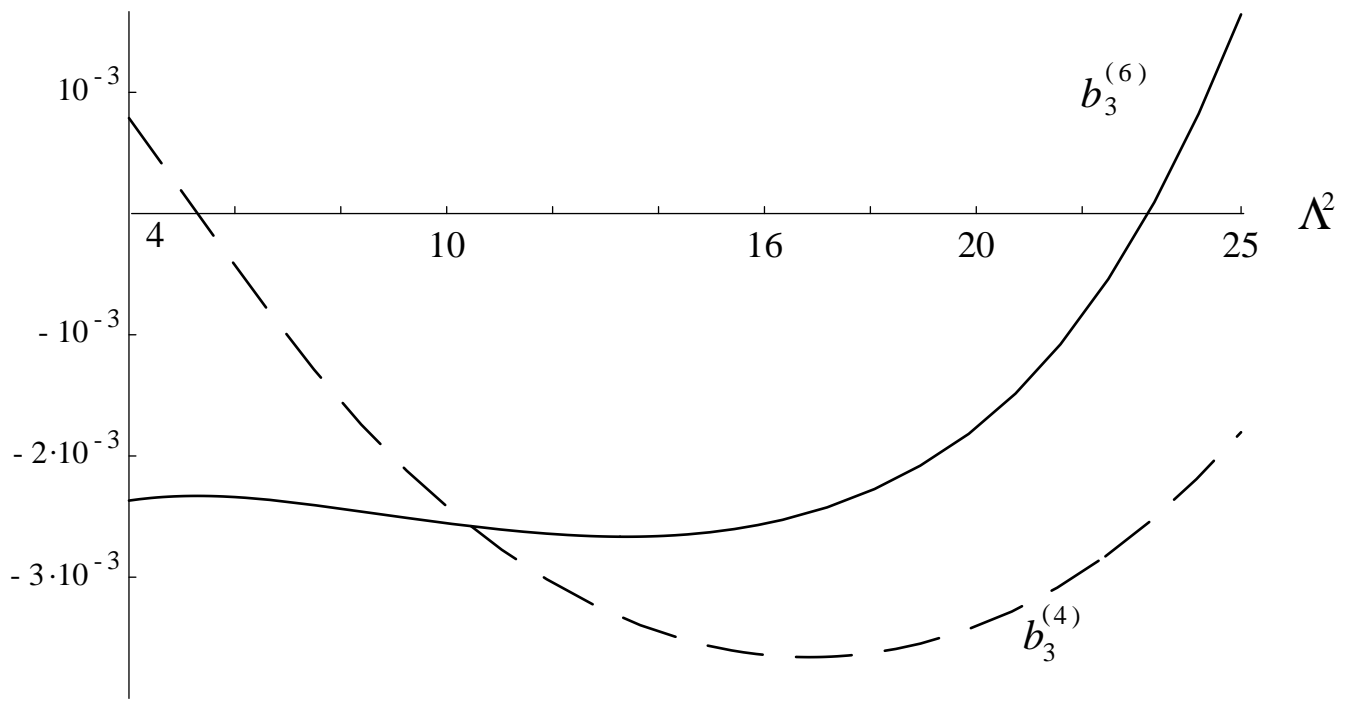
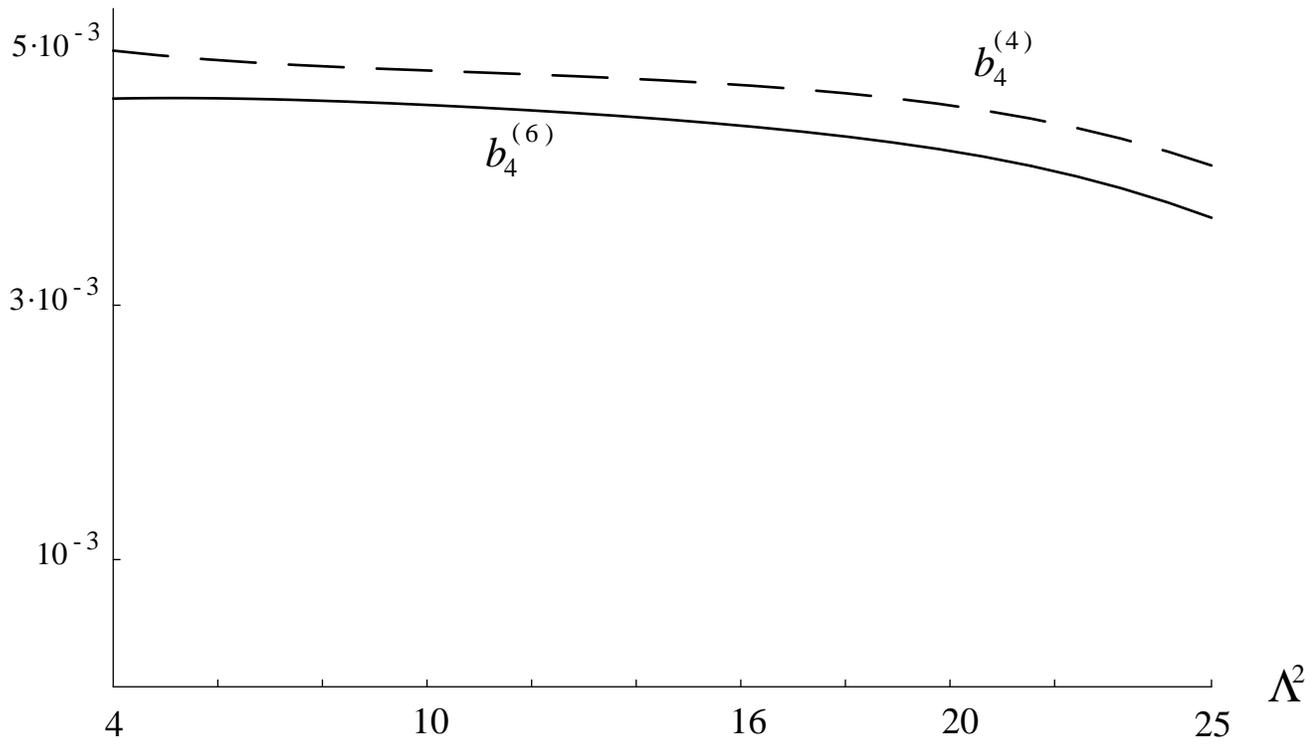
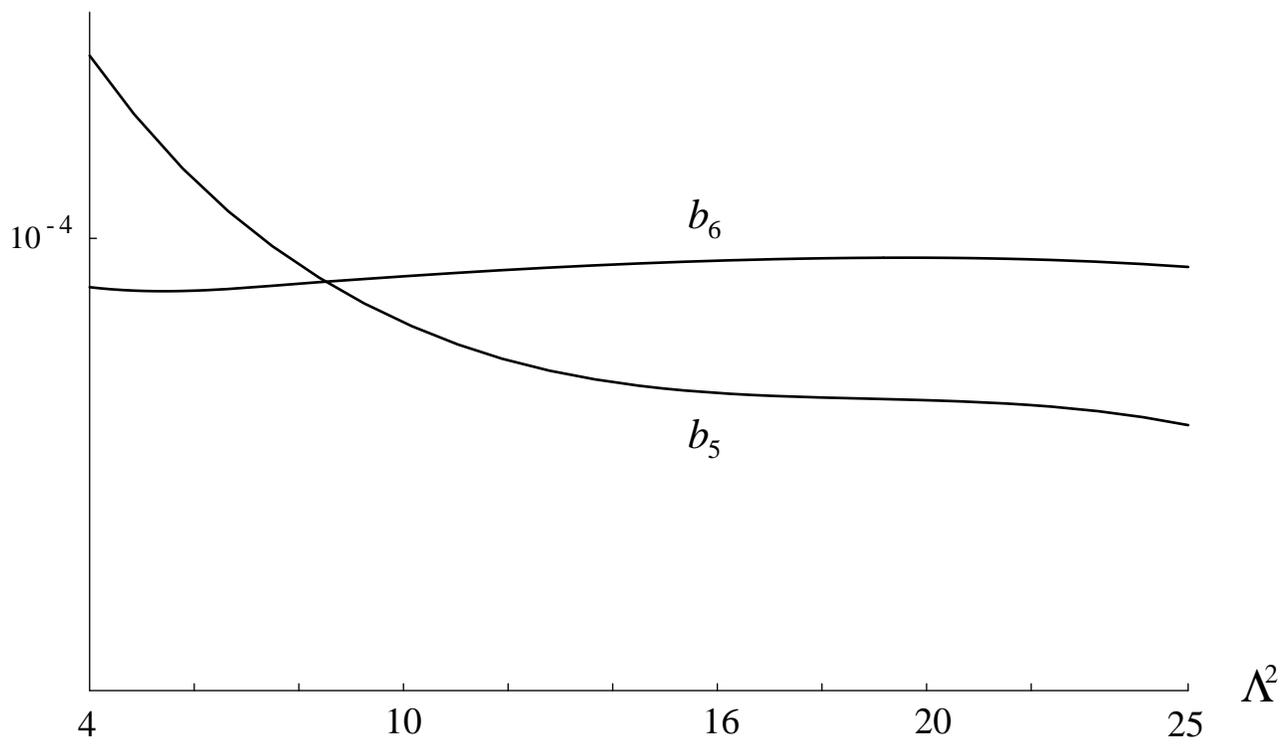
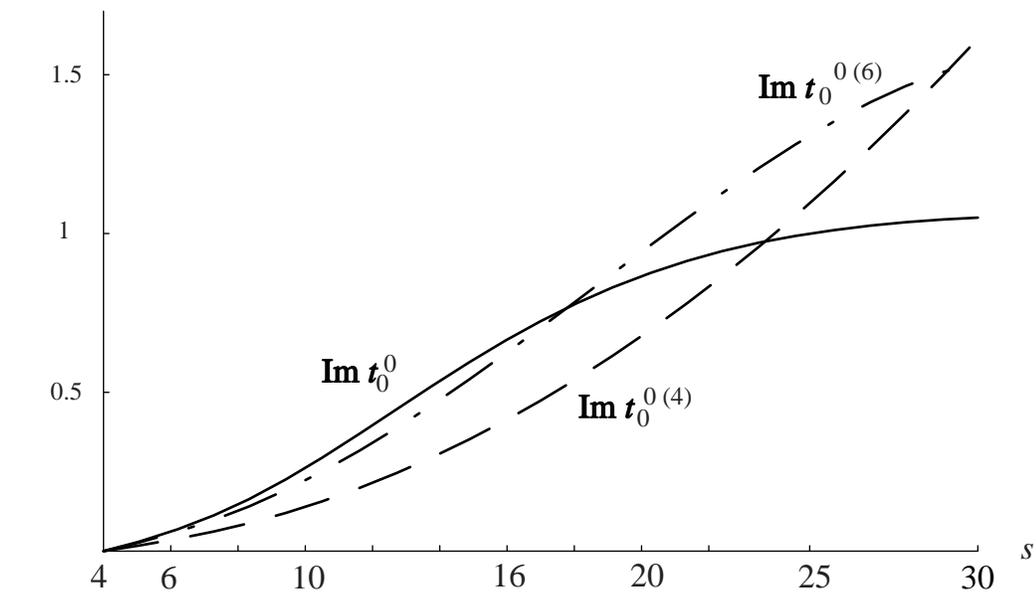
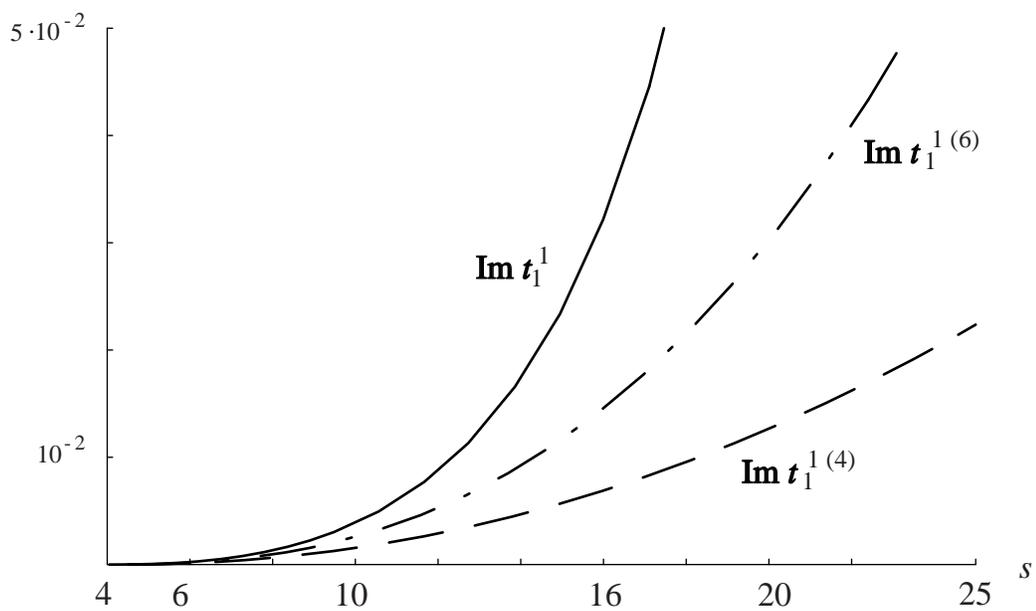
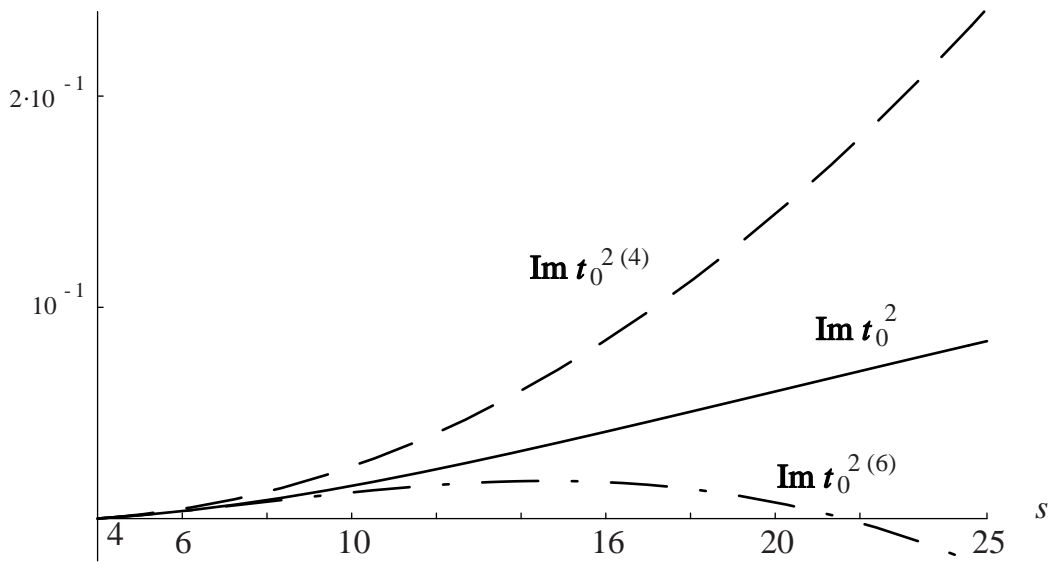
\end{document}